\documentclass[fleqn,usenatbib]{mnras}

\usepackage{newtxtext,newtxmath}

\usepackage[T1]{fontenc}
\usepackage{ae,aecompl}

\usepackage{graphicx}	
\usepackage{amsmath}	
\usepackage{amssymb}	

\title[LINERs are not RIAFs]{LIGO tells us LINERs are not optically thick RIAFs}

\author[K.E.S. Ford et al.]{K.E. Saavik Ford$^{1,2,3}$\thanks{E-mail:sford at amnh.org (SF)} and B. McKernan$^{1,2,3}$\\
$^{1}$Department of Astrophysics, American Museum of Natural History, New York, NY 10024, USA\\
$^{2}$Graduate Center, City University of New York, 365 5th Avenue, New York, NY 10016, USA\\
$^{3}$Department of Science, BMCC, City University of New York, New York, NY 10007, USA\\
}

\date{Accepted XXX. Received YYY; in original form ZZZ}

\pubyear{2015}

\begin{document}
\label{firstpage}
\pagerange{\pageref{firstpage}--\pageref{lastpage}}
\maketitle

\begin{abstract}
Low ionization emission line regions (LINERs) are a heterogeneous collection of up to $1/3$ of galactic nuclei in the local Universe. It is unclear whether LINERs are simply the result of low accretion rates onto supermassive black holes or whether they include a large number of optically thick radiatively inefficient but super-Eddington accretion flows (RIAFs). Optically thick RIAFs are typically disks of large scale-height or quasi-spherical gas flows. These should be dense enough to trap and merge a large number of the stellar mass black holes, which we expect to exist in galactic nuclei. Electromagnetic observations of photospheres of accretion flows do not allow us to break model degeneracies. However, gravitational wave observations probe the interior of accretion flows where the merger of stellar mass black holes can be greatly accelerated over the field rate. Here we show that the upper limits on the rate of black hole mergers observed with LIGO demonstrate that most LINERs cannot be optically thick RIAFs. 

\end{abstract}




\section{Introduction}
LIGO is revealing a population of merging stellar mass black holes  that is extremely numerous ($\sim 112 \rm{Gpc}^{-3} \rm{yr}^{-1}$) at the upper-end of the rate estimate  \citep{LIGO18}. Merging black holes observed so far by LIGO are generally significantly more massive than those ($5-15M_{\odot}$) observed in our own Galaxy \citep{LIGO18}. The majority of mergers  observed so far have low $\chi_{\rm eff}$, the projected component of black hole spin onto the binary orbital angular momentum. All of the above are tantalizing clues about the nature of black hole mergers in our Universe. Though the overall rate and some aspects of the observed mass distribution can be accommodated by the predictions of field binary models \citep[e.g.][]{Belczynski2010,Dominik2013,Belczynski2016}, the relatively high observed rates and masses have prompted consideration of sites where mergers may be accelerated and repeated.\\

Stellar mass black hole (BH) density in the local Universe appears greatest in our own Galactic nucleus. The observed rate of occurrence of BH X-ray binaries implies a cusp of BHs in the central parsec \citep{Hailey18,Generozov18} consistent with previous conjectures \citep{Morris93,Miralda00} and simulations \citep{Antonini14}. For a long time, it has been understood that active galactic nucleus (AGN) disks should contain a large population of embedded objects from geometric orbital coincidence and grind-down \citep{Syer91,Arty93,GT04} as well as star formation \citep{Levin07}. As a result, a promising site to generate a very high rate of BH  mergers yielding over-massive BHs detectable with LIGO are dense disks of gas around supermassive black holes \citep{McK12,McK14,Bellovary16,Bartos17,Stone17,McK17,Secunda18}. Conversely, limits that can be placed on this channel for BH mergers will allow us to restrict otherwise poorly constrained models of AGN disks.\\

Low Ionization Nuclear Emission Region galaxies (LINERs) were identified as a class of object by \citet{Heckman80}. The LINERs are galactic nuclei with spectral lines that are low ionization compared to active galactic nuclei, particularly as identified by the [\textsc{O}\textsc{iii}]/H$\beta$ line ratio \citep{Veilleux87,HFS03,Kewley06}. Approximately $\sim 1/3$ of all galactic nuclei in the local Universe can be classified as LINERs \citep{Ho2008} but it remains unclear what fraction of LINERs are powered by low accretion rate $\alpha$-disks onto supermassive black holes (SMBH) or whether some or all are powered by radiatively inefficient accretion flows (RIAFs), sometimes also known as advection dominated accretion flows (ADAFs) \citep[e.g.][]{NarayanYi94,Narayan97,Blandford99,Yuan03}. RIAFs can occur in the case of both optically thin and optically thick accretion.  Here we focus on the possibility that a large fraction of LINERs might be powered by cooler, optically thick RIAFs, like slim disks \citep{Abramowicz88,Ohsuga05}, not optically thin, hot RIAFs such as the one that might be powering Sgr A$\ast$ in our own Galaxy \citep{Broderick11}. We note, however, that Sgr A$\ast$ would not be detectable as a LINER at typical extragalactic distances.\\

If LINERs and similar objects comprise up to $1/3$ of all galaxies \citep{Ho2008}, and if LINERs are predominantly optically thick RIAFs, this accretion mode could be responsible for substantial mass growth of nuclear SMBH while doing so 'in the dark'--not substantially contributing to the ionizing flux of the local universe. While there are many plausible RIAF models (notably including advection dominated accretion flows, or ADAFs), they all share a characteristically large aspect ratio of their accretion flows. If LINERs are predominantly optically thick RIAFs then a large fraction of the population of stellar mass black holes that we expect in galactic nuclei will end up embedded in the RIAF. Gas torques on embedded objects can promote a very high rate of black hole merger within dense accretion flows \citep[e.g.][]{McK12}. It is this characteristic of optically thick RIAFs which allows us to set limits on their frequency via LIGO  measurements of the rate of stellar mass black hole binary mergers.

\begin{figure*}
\begin{center}
	\includegraphics[width=5in]{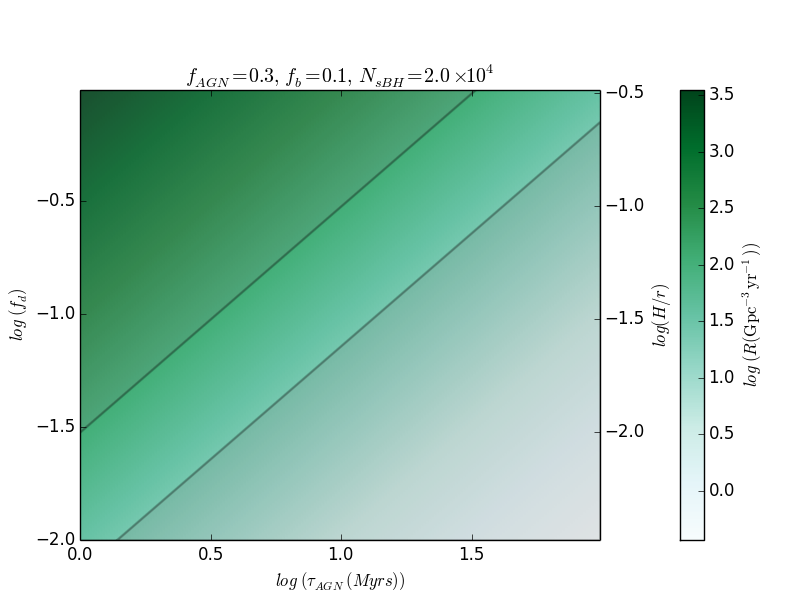}
    \caption{Fraction of nuclear stellar mass black holes embedded in an AGN disk ($f_{d}$) versus the AGN disk lifetime. Using eqn.~(\ref{eq:rate_param}), if we assume that all LINERs consist of dense, dim, optically thick, large aspect ratio ($H/r$) RIAFs, then $f_{\rm AGN} \sim 0.3$. We assume the binary fraction in the disk is $f_{b}=0.1$, the number of stellar mass black holes in galactic nuclei is $N_{\rm BH}=2 \times 10^{4}$, and $\epsilon$, a measure of the net change in the number of stellar mass black holes per galactic nucleus over an AGN duty cycle is 1 (see text). Diagonal lines correspond to the upper ($112 \rm{Gpc}^{-3} \rm{yr}^{-1}$) and lower bounds ($24 \rm{Gpc}^{-3} \rm{yr}^{-1}$) respectively, on the rate of BH mergers from \citep{LIGO18}. All LINERs must live in the space below the upper diagonal line. The color key on the RHS corresponds to the magnitude of the rate ($\cal{R}$) of BH mergers.}
    \label{fig:HoverR}
    \end{center}
\end{figure*}

\section{Rate of black hole mergers in dense, nuclear, gas disks}
From \citet{McK17} we can write the rate of black hole binary mergers in dense nuclear gas disks as
\begin{equation}
    {\cal{R}}=\frac{N_{\rm GN}N_{\rm BH} f_{\rm AGN} f_{\rm d} f_{\rm b} \epsilon}{\tau_{\rm AGN}}
\label{eq:rate}
\end{equation}
where $N_{\rm GN}$ is the number density of galactic nuclei (per $\rm{Mpc}^{-3}$), $N_{\rm BH}$ is the number of stellar mass black holes in the central $\rm{pc}^{3}$ around the supermassive black hole (SMBH), $f_{\rm AGN}$ is the fraction of galactic nuclei that are active for time $\tau_{\rm AGN}$, $f_{\rm d}$ are the fraction of nuclear stellar mass black holes that end up in the AGN disk and $\epsilon$ is the fractional change in $N_{\rm BH}$ over an AGN duty cycle. We note that this equation assumes the lifetime of a binary embedded in an AGN disk is less than $\tau_{\rm AGN}$. Binaries embedded in a gas disk should merge within $10^{5}$yrs \citep[][]{Baruteau11,McK17}, so this is a reasonable assumption (see also sec.~\ref{sec:lifetime}). Of the parameters in eqn.~(\ref{eq:rate}) only $N_{\rm GN}$ is relatively well constrained from Schecter function fits \citep[e.g.][]{Baldry12}. The other parameters in eqn.~(\ref{eq:rate}) are not very well constrained by observations, simulations or theoretical considerations and we discuss several of them below.\\

If we ignore dwarf galaxies and only count galaxies of mass $\geq 1/3$ the mass of the Milky Way, such that we count SMBH of mass $\geq 10^{6}M_{\odot}$, then $N_{\rm GN} \sim 0.006 \rm{Mpc}^{-3}$ \citep{Baldry12}. From observations of X-ray binaries around SgrA* by \citet{Hailey18}, \citet{Generozov18} extrapolate $N_{\rm BH} \sim 2 \times 10^{4}$ stellar mass BH within the central parsec of the Galaxy, which is consistent with predictions from a range of mechanisms \citep{Miralda00,Antonini14}. The binary fraction ($f_{b}$) of black holes in galactic nuclei without gas may range between $\sim 0.01-0.2$ depending on the number density of ionizing encounters \citep{Leigh18, McK17}. However, BH trapped by dense disk gas have a substantially higher binary fraction $f_{b} \sim 0.6$ as they migrate within the disk and encounter other BH \citep{Secunda18}. Thus, a fiducial overall binary fraction of $f_{b} \sim 0.1$ within a galactic nucleus containing a gas disk, seems reasonable. The parameter $f_{d}$ is the fraction of BH in the galactic nucleus that end up inside the AGN disk. To a first approximation, we can say $f_{d} \geq H/r$ the disk aspect ratio. This is because geometrically, we expect the number of orbits of a nuclear star cluster to live inside the AGN disk to correspond approximately to the disk aspect ratio ($H/r$). If we assume that the AGN disk is cylindrical, and of thickness $H/r$, then $f_{\rm d}= (H/r)$. Over time BH in the galactic nucleus with orbits oriented at an angle to the AGN disk will pass through the disk twice per orbit and experience a drag force which will tend to align the BH orbit with the plane of the disk. Thus, we expect $f_{d}> H/r$ over a long enough AGN lifetime. 

\subsection{AGN disk lifetime: $\tau_{\rm AGN}$}

\label{sec:lifetime}

AGN lifetimes are poorly constrained. A complete range of lifetimes allowed by all methods and constraints spans $\approx 10^{5}-10^{8}$yrs \citep[e.g.][]{Martini04,Graham19}. The overall quasar duty cycle (or the total amount of time spent by a galactic nucleus as a quasar) is believed to be $O(10^{8})$yrs since $f_{AGN} \sim 0.01$ for quasars, and this estimate is consistent with several modes of estimating average quasar lifetimes \citep{Martini04}. Mass doubling of SMBH via Eddington accretion occurs in $\sim 40$Myrs, so a total period of O($10^{8}$)yrs allows for approximately an order of magnitude increase in $M_{\rm SMBH}$ due to gas accretion in a Hubble time. However, this estimate of the duty cycle \emph{does not} constrain the lifetime of individual quasar episodes (i.e. $\tau_{\rm AGN}$). Furthermore, this estimate ignores more numerous, less luminous AGN phases such as the Seyfert AGN or LINERs.

Individual AGN episodes could be quite short in practice. For example, an accretion disk of mass $10^{-2}M_{\rm SMBH}$ will be consumed in only $\sim$ few Myrs at the Eddington rate. Stochastic accretion episodes are expected to arrive from random directions in the galactic bulge yielding $\tau_{\rm AGN} \sim 10^{5-6}$yr \citep[e.g.][]{King15}. Observations of large samples of AGN over relatively long time baselines enable estimates of the turn-on/turn-off rate, which do help constrain  $\tau_{\rm AGN} > 10^{5}$yr \citep[since][find O(1 in 10$^5$)AGN change state each year]{Graham19}. However, such short-lived episodes may run together if there is a large fuel source (a dusty torus) subject to instabilities near a SMBH. The running together of short-lived accretion episodes, may account for the duration of long-lived Mpc-scale radio jets. If AGN disks are not to collapse in a burst of star-formation under their own self-gravity beyond about $\sim 10^{3}r_{g}$, where $r_{g}=GM_{\rm SMBH}/c^{2}$, some additional source of heating of the outer disk is required  \citep[e.g.][]{SG03}. If no such heating were available, the innermost disk would be a mere $\sim 10^{3}r_{g}$ in size, a size estimate contradicted by numerous lines of observational evidence, including the SED \citep[e.g.][]{SG03}, reverberation mapping \citep[e.g.][]{Yu18}, and maser mapping\citep[e.g.][]{Zhao18}. The viscous timescale at a given radius ($t_{\nu}$) can be parameterized as \citep{Stern18}
\begin{equation}
    t_{\nu} \approx 10^{4}{\rm yr} \left(\frac{H/r}{0.04}\right)^{-2} \left( \frac{\alpha}{0.03}\right)^{-1} \left( \frac{M_{\rm SMBH}}{10^{8}M_{\odot}}\right) \left( \frac{R}{10^{3}r_{g}}\right)^{3/2}
\end{equation}
where $\alpha$ is the disk viscosity parameter and a truncated AGN disk is short lived. Such very short timescales for individual AGN episodes may be consistent with the O($10^{5}$yr) duration of the response of circum-galactic medium to past quasar episodes \citep{Schaw15}. However, if additional heating stabilizes the outer disk to $\sim 10^{5}r_{g} \sim 1{\rm pc}$ then $t_{\nu} \approx 10$Myr, though shorter lifetimes are possible for modestly larger $\alpha$ and $H/r$. Additionally, a series of closely spaced, short-lived episodes of activity, followed by a longer, fully quiescent period may be dynamically indistinguishable from a single, modest-length, continuous period of activity--as long as the brief quiescent periods are short compared to the dynamical relaxation time of the nuclear star cluster (O(100Myr)). 

In summary, there are many theoretical reasons to expect both short-lived and long-lived AGN disks. The observational constraints may conflict, and are challenging to obtain. Quite possibly the AGN phenomenon intrinsically spans a wide range of disk lifetimes. Thus, an alternative method for constraining AGN disk lifetimes would be an important contribution to our understanding of the phenomenon.

\subsection{AGN disk aspect ratio: $H/r$}

AGN disk thickness as a function of radius is given by $H={c_{s}}/{\Omega}$
where $c_{s}$ is the sound speed in the disk gas at that radius and $\Omega=(GM_{\rm SMBH}/r^{3})^{1/2}$ is the Keplerian orbital frequency. Writing $r$ in terms of the gravitational radius of the SMBH, $r_{g}=GM_{\rm SMBH}/c^{2}$ we can write
\begin{equation}
    \frac{H}{r} \sim \left( \frac{c_{s}}{c}\right) \left(\frac{r}{r_{g}}\right)^{1/2}
\end{equation}
so the AGN disk aspect ratio simply depends on the ratio ($c_{s}/c$) in the gas. The sound speed in gas is given by $c_{s}^{2}={P_{\rm total}}/{\rho}$
where $P_{\rm total}=P_{\rm gas}+P_{\rm rad}+P_{\rm mag}$ is the total pressure in the gas of density $\rho$ at that radius and
\begin{eqnarray}
    P_{\rm gas} &=&\frac{\rho k_{B} T}{m_{H}} \\
    P_{\rm rad} &=&\left(\frac{\tau}{c}\right) \sigma T_{\rm eff}^{4} \\
    P_{\rm mag} &=& \frac{B^{2}}{2\mu_{0}}
\end{eqnarray}
where  $T$ is the midplane temperature, $T_{\rm eff}$ is the effective temperature of the disk photosphere, $\tau$ is the optical depth from the midplane to the photosphere, $B$ is the magnetic field strength and the constants have the usual meanings. A standard thin disk model has a disk mid-plane temperature profile which goes as $T(r)=T_{\rm ISCO} r^{-3/4}$ where $T_{\rm ISCO}$ is the temperature at the innermost stable circular orbit (ISCO). Thus, close to the ISCO, $H/r$ is large. However, at distances $\sim 10^{3}r_{g}$ from the ISCO, where the disk is cool and dense and therefore $c_{s}$ is small, the disk aspect ratio tends to be small $H/r \sim 10^{-2},10^{-3}$ \citep{SG03,Thompson05}. At very large distances, $H/r$ can increase again. 

However, a $r^{-3/4}$ dependence of the disk temperature profile assumes no irradiation of the outer disk and ignores the impact on the disk of multiple embedded objects. These embedded objects will each accrete, migrate and collide, which will contribute significant local disk heating at large radii. Localized disk heating will tend to smear out into annuli over an orbital time. Disk-crossing orbiters will also contribute significant disk heating as they pass through the disk. Thus, allowing for a population of embedded and disk-crossing objects is likely to increase $H/r$ significantly over the gas-only radial temperature profile \citep{McK12,McK14}.

The above discussion applies to accretion disks which are relatively thin due to the efficiency of radiative cooling. In a disk where the photon diffusion timescale ($t_{\gamma}$) is less than $t_{\nu}$, the viscous (accretion) timescale, the disk becomes a radiatively inefficient accretion flow (RIAF) and can be geometrically very thick ($H/r \sim 0.1-1$) \citep[e.g.][]{NarayanYi94,Narayan97,Blandford99,Yuan03}.

\subsection{Black hole mergers over an AGN duty cycle}
\label{sec:duty}
The parameter $\epsilon$ in eqn.~(\ref{eq:rate}) is a way of estimating the overall change in BH number in a galactic nucleus over a full AGN cycle. If the AGN lasts for some time ($\tau_{\rm AGN}$) and the quiescent period before the next AGN episode is $t_{\rm Q} \gg \tau_{AGN}$ then the effective AGN duty cycle is $f_{\rm AGN} = \tau_{\rm AGN}/(\tau_{\rm AGN}+ t_{\rm Q})$. We can write $\epsilon$ as the number of BH after one AGN duty cycle divided by the initial number of BH, or
\begin{equation}
    \epsilon = \frac{N_{\rm BH} (\tau_{\rm AGN} + t_{\rm Q})}{N_{\rm BH}(0)}
\end{equation}
which can be written as
\begin{equation}
    \epsilon= 1 -\frac{\tau_{\rm AGN}{\cal{R}}_{\rm AGN} }{N_{\rm BH}(0)} + \frac{(\tau_{\rm AGN} + t_{\rm Q})\dot{N}_{\rm BH} }{N_{\rm BH}(0)}
\end{equation}
where ${\cal{R}}_{\rm AGN}$ is the mean rate of BH merger per AGN. The quantity $\dot{N}_{\rm BH}=N^{+}_{\rm BH}-N^{-}_{\rm BH}$ corresponds to the number of BH ($N^{+}_{\rm BH}$) that have entered the central $\rm{pc}^{-3}$, via star formation and dynamical friction over the AGN duty cycle, minus the number of BH ($N^{-}_{\rm BH}$) that have left the central $\rm{pc}^{-3}$ over the same time, via scattering or orbital evolution via energy equipartition. The parameter $\epsilon \geq 1$ if
\begin{equation}
    \dot{N}_{\rm BH} \geq f_{\rm AGN}{\cal{R}}_{\rm AGN}.
\end{equation}
Given that $N_{\rm GN} \sim 6 \times 10^{-3} \rm{Mpc}^{-3}$ and ${\cal{R}} <112 \rm{Gpc}^{-3} \rm{yr}^{-1}$ then ${\cal{R}}_{\rm AGN} \leq 20 \rm{Myr}^{-1}/ \rm{f}_{\rm AGN}$. So, for $\epsilon \geq 1$ we require $\dot{N}_{BH} \geq 20 \rm{Myr}^{-1}$ which seems a modest and plausible requirement given the star formation rates inferred from observations \citep{Hopkins2010} and dynamical friction estimates \citep{Morris93,Antonini12}. So, at a minimum we can expect $\epsilon \sim 1$.

\subsection{Merger rate as a constraint on disk models}
\label{sec:constraints}
Based on the preceding arguments we can parameterize the rate of binary BH mergers in AGN as 
\begin{eqnarray}
    {\cal{R}}& \approx &1.2 \times 10^{4}\rm{Gpc}^{_3} \rm{yr}^{-1} f_{\rm AGN} f_{\rm d} \nonumber \\  
    &\times & \left(\frac{N_{\rm GN}}{0.006 \rm{Mpc}^{-3}}\right) \left(\frac{N_{\rm BH}}{2 \times 10^{4}}\right)\left(\frac{f_{\rm b}}{0.1}\right) \left(\frac{\tau_{\rm AGN}}{1\rm{Myr}} \right)^{-1} \left(\frac{\epsilon}{1}\right).
\label{eq:rate_param}
\end{eqnarray}
\citet{LIGO18} find before O3 that the rate of binary black hole mergers in the local Universe is $53.2^{+58.5}_{-28.8}$ at 90$\%$ confidence. We can immediately see that the rate in eqn.~(\ref{eq:rate_param}) could be orders of magnitude too large, even if AGN disks are the {\it only} site of LIGO-detected sBBH mergers. The problem becomes worse if we include the likely possibility that there exist other channels (including field binaries) contributing to the observed rate.
Now we can vary the parameters $f_{\rm AGN}$ and $f_{\rm d} \propto (H/r)$ over a range of possible $\tau_{\rm AGN}$ in order to quantify the constraints on AGN model parameters set by the LIGO rate measurements.

\section{Results}
By assuming that all LINERs are dense, optically thick RIAFs, we can immediately assume $f_{\rm AGN} \sim 0.3$ in eqn.~(\ref{eq:rate_param}) above. Implicitly, we can also assume $H/r >0.1$, possibly as large as a quasi-spherical $H/r \sim 1$ \citep{Abramowicz88}. We can also assume that the same analysis that applies to regular, dense AGN disks \citep{McK12,McK14} must also apply to LINERs. That is, we can expect efficient orbital grind-down, migration and merger due to gas torques on embedded stellar mass BH.\\

Fig.~\ref{fig:HoverR} shows $f_{d}$ versus $\tau_{\rm AGN}$ using  eqn.~(\ref{eq:rate_param}) and the above assumptions. The diagonal lines on the plot correspond to the upper ($112 \rm{Gpc}^{-3} \rm{yr}^{-1}$) and lower bounds ($24 \rm{Gpc}^{-3} \rm{yr}^{-1}$) respectively, on the rate of BH mergers from \citep{LIGO18}. So all AGN/LINERs must live in the space below the upper diagonal line. The color key on the RHS corresponds to the magnitude of the rate ($\cal{R}$) of BH mergers.\\

The important point from Fig.~\ref{fig:HoverR} is that we must live in a Universe below the upper diagonal line. If AGN are responsible for {\it all} LIGO detected mergers, we may, at most, lie {\it on} the line. If on the other hand, AGN are responsible for only a fraction of LIGO detected mergers, we must lie somewhere below the diagonal line, possibly substantially below it.
Even with the current rate measurements, we must exclude the possibility of LINERs typically hosting 'fat', dense accretion flows, unless those flows persist for lifetimes $>5$Myr. However, if LINERs were to persist for such long lifetimes and account for $\sim 1/3$ of all galactic nuclei, the resulting  (super-Eddington) rate of black hole growth \citep{Abramowicz88,Ohsuga05} in these systems would quickly exceed the measurements of total local SMBH mass. Therefore, we conclude that LIGO limits on the rate of black hole mergers rules out the possibility that LINERs consist mostly of cool, optically thick accretion flows. We note that we cannot exclude the possibility that a small subset of LINERs do host cool, optically thick RIAFs. We note that the values we have chosen for $f_{b}$ and $N_{BH}$, while reasonable, could have values other than those specified. However, given other constraints ($f_{\rm AGN}=0.3$, $\tau_{\rm AGN}<5$Myr), our conclusions apply as long as the product $f_{b}N_{BH}\geq 350$.\\

This same strategy can be employed to constrain many difficult to measure parameters of AGNs. These constraints become even more stringent if we can independently measure the AGN contribution to the LIGO rate. This may be possible in LIGO O3 by studying the distribution of BH merger mass ratios and the $\chi_{\rm eff}$ distribution of the merging BHs \citep{Fishbach17,Gerosa17,McK19}. 
However, in the next few years, with planned upgrades yielding BBH merger detection rates as high as O($10^3$ yr$^{-1}$), we will definitely be able to independently constrain the fractional AGN contribution to the measured LIGO rate.
This relies on a statistical strategy first proposed by \citet{Bartosstats2017}; for AGN fractional contributions $>0.3$, given typical space densities of AGN, we can measure their contribution simply by measuring the number of AGN per LIGO error volume, and comparing that to the expected space density. If AGN are the source of at least 30\% of LIGO mergers, an excess of AGN will be detectable with fewer than 1000 detected mergers, assuming typical sky localization and no improvements to current AGN catalogs. \\

\section{Conclusions}
The upper bound to the rate of black hole binary mergers observed by LIGO ($\sim 112 \rm{Gpc}^{-3} \rm{yr}^{-1}$) tells us that LINERs are mostly not cool, optically thick RIAFs or slim disks, with super-Eddington accretion. \footnote{Our conclusion also rests on the assumption that some BBH mergers occur in AGN disks--though given the expected overabundance of stellar mass black holes in galactic nuclei, the absence of such mergers would seem nearly pathological. This assumption is further supported by recent work concluding that at least one LIGO detected merger did occur in a galactic nucleus or AGN disk \citep{Gerosa19,Yang19}.} Instead, LINERs must be predominantly powered by low accretion rate $\alpha$disks, or optically thin RIAFs or some combination thereof. Future improvements in the precision of the measured rate will enable us to provide further constraints on various parameters of AGN disks; independent measurements of the AGN contribution to the sBBH merger rate, and the characteristics of those mergers \citep[e.g.][]{Fishbach17,Gerosa17,Bartosstats2017,McK19} will further constrain disk models. Importantly, this technique has the capacity to observationally constrain AGN disk lifetimes--a quantity that is otherwise extremely challenging to measure.
Finally, we note this is a clear demonstration of the power of gravitational wave astronomy to open new and unexpected astrophysical windows.
\\

\section*{Acknowledgements}
KESF \& BM are supported by NSF grant 1831412 and BMCC Faculty Development grants. We acknowledge very useful discussions with Jillian Bellovary, Cole Miller, Yuri Levin, Mordecai-Mark MacLow, Richard O'Shaughnessy, Avi Loeb, Phil Armitage, Chiara Mingarelli, and Will Farr. We would also like to thank the participants at the Black Holes in the Disks of Active Galactic Nuclei Workshop, March 11-13, 2019, and especially the Center for Computational Astrophysics at The Flatiron Institute for sponsoring this workshop.





\begin{thebibliography}{99}
\bibitem[\protect\citeauthoryear{Abramowicz et al.}{1988}]{Abramowicz88} Abramowicz, M. A., Czerny, B., Lasota, J. P. \& Szuszkiewicz, E., 1988, ApJ, 332, 646
\bibitem[\protect\citeauthoryear{Antonini}{2012}]{Antonini12} Antonini, F. \& Merritt D., 2012, ApJ, 745, 83
\bibitem[\protect\citeauthoryear{Antonini}{2014}]{Antonini14} Antonini, F. 2014, ApJ, 794, 106
\bibitem[\protect\citeauthoryear{Artymowicz}{1993}]{Arty93} Artymowicz, P. 1993, ApJ, 409, 592
\bibitem[\protect\citeauthoryear{Baldry et al.}{2012}]{Baldry12} Baldry I. K. et al., 2012, MNRAS, 421, 621
\bibitem[\protect\citeauthoryear{Bartos et al.}{2017}]{Bartosstats2017}
Bartos, I., Haiman, Z., Marka, Z., Metzger, B.~D., Stone, N.~C. \& Marka, S., 2017, Nature Communications, 8, 831
\bibitem[\protect\citeauthoryear{Bartos et al.}{2017}]{Bartos17} Bartos I. et al., 2017, ApJ, 835, 165
\bibitem[\protect\citeauthoryear{Baruteau et al.}{2011}]{Baruteau11} Baruteau C., Cuadra J., Lin D.N.C., 2011, ApJ, 726, 28
\bibitem[\protect\citeauthoryear{Belczynski et al.}{2010}]{Belczynski2010} Belczynski, K., Dominik, M., Bulik, T., O'Shaughnessy, R., Fryer, C., \& Holz, D.~E., 2010, ApJ, 715, L138
\bibitem[\protect\citeauthoryear{Belczynski et al.}{2016}]{Belczynski2016}
Belczynski, K., Holz, D.~E., Bulik, T., \& O'Shaughnessy, R., 2016, Nature, 534, 512
\bibitem[\protect\citeauthoryear{Bellovary et al.}{2016}]{Bellovary16} Bellovary J. et al., 2016, ApJ, 819, L17
\bibitem[\protect\citeauthoryear{Blandford \& Begelman}{1999}]{Blandford99} Blandford R.D. \& Begelman M.C., 1999, MNRAS, 303, L1
\bibitem[\protect\citeauthoryear{Broderick et al.}{2011}]{Broderick11} Broderick A.E., Fish V.L., Doeleman S.S. \& Loeb A., 2011, ApJ, 735, 110

\bibitem[\protect\citeauthoryear{Dominik et al.}{2013}]{Dominik2013} Dominik, M., Belczynski, K., Fryer, C., Holz, D.~E., Berti, E., Bulik, T., Mandel, I. \& O'Shaughnessy, R., 2013, ApJ, 779, 72
\bibitem[\protect\citeauthoryear{Fishbach et al.}{2017}]{Fishbach17} Fishbach M., Holz D.E. \& Farr B., 2017, ApJ, 840, L24
\bibitem[\protect\citeauthoryear{Generozov et al.}{2018}]{Generozov18} Generozov A., Stone N.C., Metzger B.D. \& Ostriker J.P., 2018, MNRAS, 478, 4030
\bibitem[\protect\citeauthoryear{Gerosa \& Berti}{2017}]{Gerosa17} Gerosa D. \& Berti E., 2017, PhRvD, 95, 124046
\bibitem[\protect\citeauthoryear{Gerosa \& Berti}{2019}]{Gerosa19} Gerosa D. \& Berti E., 2019, PhRvD (submitted), arXiv:1906.05295
\bibitem[\protect\citeauthoryear{Goodman \& Tan}{2004}]{GT04} Goodman J. \& Tan J.C., 2004, ApJ, 608, 108
\bibitem[\protect\citeauthoryear{Graham et al.}{2019}]{Graham19} Graham M.J. et al., 2019, MNRAS (submitted), arXiv:1905.02262
\bibitem[\protect\citeauthoryear{Hailey et al.}{2018}]{Hailey18} Hailey C.J. et al., 2018, Nature, 556, 70
\bibitem[\protect\citeauthoryear{Heckman}{1980}]{Heckman80} Heckman T.M., 1980, A\&A, 87, 152
\bibitem[\protect\citeauthoryear{Ho, Filippenko \& Sargent}{2003}]{HFS03} Ho L.C., Filippenko A.V. \& Sargent W., 2003, ApJ, 583, 159
\bibitem[\protect\citeauthoryear{Ho}{2008}]{Ho2008} Ho L.C., 2008, ARA\&A, 46, 475
\bibitem[\protect\citeauthoryear{Hopkins}{2010}]{Hopkins2010} Hopkins P.F. et al., 2010, MNRAS, 402, 1693
\bibitem[\protect\citeauthoryear{Kewley et al.}{2006}]{Kewley06} Kewley L.J., Groves B., Kauffmann G. \& Heckman T.M., 2006, MNRAS, 372, 961
\bibitem[\protect\citeauthoryear{King \& Nixon}{2015}]{King15} King A.  \& Nixon C.J., 2015, MNRAS, 453, L46
\bibitem[\protect\citeauthoryear{Leigh}{2018}]{Leigh18} Leigh N.W.C. et al., 2018, MNRAS, 474, 5672
\bibitem[\protect\citeauthoryear{Levin}{2007}]{Levin07} Levin Y., 2007, MNRAS, 374, 575L
\bibitem[\protect\citeauthoryear{LIGO \& VIRGO}{2018}]{LIGO18} LIGO \& VIRGO Scientific Collaborations, 2018, ApJ (submitted), arXiv:1811.12940
\bibitem[\protect\citeauthoryear{Martini}{2004}]{Martini04} Martini P., 2004, arXiv:astro-ph/0304009
\bibitem[\protect\citeauthoryear{McKernan et al.}{2012}]{McK12} McKernan B. et al., 2012, MNRAS, 425, 460
\bibitem[\protect\citeauthoryear{McKernan et al.}{2013}]{McK13} McKernan B. et al., 2013, 432, 168
\bibitem[\protect\citeauthoryear{McKernan et al.}{2014}]{McK14} McKernan B. et al., 2014, 441, 900
\bibitem[\protect\citeauthoryear{McKernan et al.}{2015}]{McK15} McKernan B. et al., 2015, 452, L1
\bibitem[\protect\citeauthoryear{McKernan et al.}{2018}]{McK17} McKernan B. et al., 2018, ApJ, 866, 66
\bibitem[\protect\citeauthoryear{McKernan et al.}{2019}]{McK19} McKernan B., Ford K.E.S., O'Shaughnessy R. \& Wysocki D., 2019, MNRAS (submitted), arXiv: TBD
\bibitem[\protect\citeauthoryear{Miralda-Escud\'{e} \& Gould}{2000}]{Miralda00} Miralda-Escud\'{e} J. \& Gould A., 2000, ApJ, 545, 847
\bibitem[\protect\citeauthoryear{Morris}{1993}]{Morris93} Morris M., 1993, ApJ, 408, 496
\bibitem[\protect\citeauthoryear{Narayan \& Yi}{1994}]{NarayanYi94} Narayan R. \& Yi I., 1994, ApJ, 428, L13
\bibitem[\protect\citeauthoryear{Narayan et al.}{1997}]{Narayan97} Narayan R., Kato S. \& Honma F., 1997, ApJ, 476, 49
\bibitem[\protect\citeauthoryear{Ohsuga et al.}{2005}]{Ohsuga05} 
Ohsuga, K., Mori, M., Nakamoto, T. \& Mineshige, S., 2005, ApJ, 628, 368
\bibitem[\protect\citeauthoryear{Schawinski et al.}{2015}]{Schaw15} Schawinski K. et al., 2015, ApJ, 451, 2517
\bibitem[\protect\citeauthoryear{Secunda et al.}{2018}]{Secunda18} Secunda A. et al., 2018, ApJ, submitted, arXiv:1807.02859
\bibitem[\protect\citeauthoryear{Sigurdsson \& Phinney}{1993}]{SigPhin} Sigurdsson S. \& Phinney E.S., 1993, ApJ, 415, 631
\bibitem[\protect\citeauthoryear{Sirko \& Goodman}{2003}]{SG03} Sirko E. \& Goodman J., 2003, MNRAS, 341, 501
\bibitem[\protect\citeauthoryear{Stern et al.}{2018}]{Stern18} Stern D. et al., 2018, ApJ, 864, 27
\bibitem[\protect\citeauthoryear{Stone et al.}{2017}]{Stone17} Stone N.C. et al., 2017, ApJ, 464, 946
\bibitem[\protect\citeauthoryear{Syer et al.}{1991}]{Syer91} Syer D., Clarke C. \& Rees M.J., 1991, MNRAS, 250, 505
\bibitem[\protect\citeauthoryear{Thompson et al.}{2005}]{Thompson05} Thompson T., Quataert E. \& Murray N., 2005, ApJ, 630, 167
\bibitem[\protect\citeauthoryear{Veilleux \& Osterbrock}{1987}]{Veilleux87} Veilleux S. \& Osterbrock D.E., 1987, ApJS, 63, 295
\bibitem[\protect\citeauthoryear{Yang et al.}{2019}]{Yang19} Yang Y. et al., 2019, PRL (submitted), arXiv:1906.09281
\bibitem[\protect\citeauthoryear{Yu et al.}{2018}]{Yu18} Yu Z. et al., 2018, ApJ (submitted), arXiv:1905.02262
\bibitem[\protect\citeauthoryear{Yuan et al.}{2003}]{Yuan03} Yuan F., Quataert E. \& Narayan R., 2003, ApJ, 598, 301
\bibitem[\protect\citeauthoryear{Zhao et al.}{2018}]{Zhao18} Zhao W. et al., 2018, ApJ, 854, 124
\end{thebibliography}





\bsp	
\label{lastpage}
\end{document}